\documentclass[12pt]{article}

\usepackage{epsfig}
\usepackage{amsmath}
\usepackage{soul,color}
\usepackage{bm}
\usepackage{epsfig}
\usepackage{natbib}
\newcommand{\tr}{^{\prime}}
\def\b#1{\mbox{\boldmath $#1$}}    
\def\bl#1{\mbox{\footnotesize \boldmath {$#1$}}} 
\def\cg#1{\mbox{${\cal #1}$}}      

\newcommand{\logit}{\mbox{logit }} 
\renewcommand{\th}{\theta}

\newcommand{\al}{\alpha}
\newcommand{\be}{\beta}

\newcommand{\ga}{\gamma}

\def\bTheta{\mbox{\boldmath$\Theta$}}
\def\btheta{\mbox{\boldmath$\theta$}}

\def\boeta{\mbox{\boldmath$\eta$}}

\usepackage{bm}
\usepackage{epsfig}

\begin{document}
\title{\vspace*{-2cm}Joint Assessment of the Differential Item Functioning \\ and
Latent Trait Dimensionality\\ of  Students' National Tests}
\author{Michela Gnaldi$^*$, Francesco Bartolucci$^*$,
Silvia Bacci\footnote{Department of Economics, Finance and
Statistics, University of Perugia, Via A. Pascoli, 20, 06123
Perugia. E--mail: gnaldi@stat.unipg.it, bart@stat.unipg.it,
sbacci@stat.unipg.it}}
\date{}
\maketitle
\vspace*{-0.5cm}
\begin{abstract}
Within the educational context, students' assessment tests are routinely validated
through Item Response Theory (IRT) models which assume unidimensionality and absence of Differential Item Functioning (DIF).
In this paper, we investigate if such assumptions hold for two national tests
administered
in Italy to middle school students in June 2009: the Italian Test and
the Mathematics Test. To this aim, we rely on an extended class of
multidimensional latent class IRT models characterised by:
({\em i}) a two-parameter logistic parameterisation for the conditional
probability of a correct response, ({\em ii}) latent traits represented
through a random vector with a discrete distribution, and ({\em iii}) the
inclusion of (uniform) DIF to account for
students' gender and geographical area. A classification of the items into
unidimensional groups is also proposed and represented
by a dendrogram, which is obtained from a hierarchical clustering algorithm.
The results provide evidence for DIF effects for both Tests. Besides, the
assumption of unidimensionality is strongly rejected for the Italian Test,
whereas
it is reasonable
for the Mathematics Test.

\noindent \vskip3mm \noindent {\sc Keywords:}
EM algorithm; Hierarchical clustering; Item Response Theory;
Multidimensional latent variable models; Two-parameter logistic parameterisation.
\end{abstract}

\section{Introduction}
Italian National Tests for the assessment of primary, lower middle,
and high-school students are developed and yearly collected by the
National Institute for the Evaluation of the Education System
(INVALSI). Before administration, national tests are validated
through pretesting sessions. These preliminary data are analysed by
standard Classical Test and Item Response Theory (IRT) models
\citep{ham:85}.

In this paper, we focus on the Tests administered to middle school
students as they are having an increasing relevance in the Italian
education context and their collection will become compulsory in the
near future. In particular, we aim at studying if the assumptions of
the IRT models used by the INVALSI to calibrate the national Tests
are met for the ``live'' data collected by this Institution in June
2009, focusing in particular on the assumptions of unidimensionality
and of no Differential Item Functioning (DIF). The data are based on
a nationally representative sample of 27,592 students within 1,305
schools (one class is sampled in each school) and refer to students'
performances in two national tests, the Italian Test and the
Mathematics Test, administered in June 2009.

In accordance with the assumption of unidimensionality, which
characterizes the most common IRT models, responses to a set of
items only depend on a single latent trait which, in the educational
setting, can be interpreted as the student's ability. However, if
unidimensionality is not met, summarizing students' performances
through a single score, on the basis of a unidimensional IRT model,
may be misleading as test items indeed measure more than one
ability. Absence of DIF means that the items have the same
difficulty for all subjects and, therefore, difficulty does not vary
among different groups defined, for instance, by gender or
geographical area.

In connection with the Rasch model, the hypothesis of
unidimensionality  has been extensively tested in the literature on
the subject \citep{rasch:61, glas:ver:95, ver:01}. One of the main
contributions has been developed by \cite{mar:73}, who proposed to
test the hypothesis that the Rasch model holds for the whole set of
items against the hypothesis that this model holds for two disjoint
subsets of items defined in advance. Therefore, most statistical
tests proposed in the literature are based on the assumptions that:
({\em i}) item discrimination power is constant and ({\em ii}) the
conditional probability to answer a given item correctly does not
vary across different groups. It is plausible that, given the
complexity of the INVALSI study, these assumptions are not met for
the INVALSI Test items as they may not discriminate equally well
among subjects and may exhibit differential item functioning (DIF).

In line with the above issues, we illustrate an extension of the class of
multidimensional latent class IRT models developed by
\cite{bart:07} to include DIF effects. Specifically, we consider
the version of these models based on a two-parameter logistic (2PL)
parameterisation \citep{bir:68} for the conditional probability of a correct
response. The applied models are of latent class type, as they rely on the
assumption that the population under study is made up by a finite number of
classes, with subjects in the same class having the same ability level \citep
{laza:henr:68,form:95,lind:91}. Representing the ability distribution through a
discrete latent variable is more flexible than representing it by means of a
continuous distribution and is compatible with the assumption of
multidimensionality, which means that the adopted questionnaire indeed measures
more than one type of ability or dimension \citep{laza:henr:68, form:95, lind:91}.

On the basis of the extended class of models described above, we
analyse the 2009 INVALSI ``live'' data. These data are collected by
two National Tests, which are developed to assess a number of
different abilities, such as the ability to make sense of written
texts, the ability to understand expressions and equations, and so
on. As already mentioned, these Tests are of particular relevance in
the Italian educational system; moreover, their reliability is
nowadays deeply discussed. With reference to these data, in
particular, we test the hypothesis of unidimensionality and that of
absence of DIF. Moreover, we provide a clustering of the items, so
that the items in the same group are referred to the same ability.
This is obtained by performing a sequence of Wald tests between
nested multidimensional IRT models belonging to the proposed class.
The results of this clustering procedure may be effectively
illustrated by dendrograms.

The remainder of this paper is organized as follows. In the next section we
describe the INVALSI data used in our analysis. The statistical methodological
approach employed to investigate the structure of the questionnaires is described
in Section \ref{sec:methods}. Firstly, we recall the basics for the model adopted
in our study \citep{bart:07};
then we show how it can be extended to take into account DIF effects. Details
about the estimation algorithm and the use of these models to test
unidimensionaly and absence of DIF are given in Section \ref{sec:ML}.
Finally, in Section \ref{sec:appl}, we illustrate the main results obtained by
applying the proposed
approach to the INVALSI dataset and in Section \ref{sec:conc} we draw the main
conclusions of the study.

\section{The 2009 INVALSI Tests}

In 2009, the INVALSI Italian Test included two sections, a Reading Comprehension
section and a Grammar section. The first
section is based on
two texts: a narrative type text (where readers engage with imagined events and
actions) and an informational text (where readers engage with real settings);
see \cite{INV:091}.
The comprehension processes are measured by 30 items, which require
students to demonstrate a range of abilities and skills in constructing meaning
from the two written texts. Two main types of comprehension processes were
considered in developing the items: Lexical Competency, which covers the ability to make
sense of worlds in the text and to recognize meaning connections among them, and
Textual
Competency, which relates to the ability to: ({\em i}) retrieve or locate
information in
the text, ({\em ii}) make inferences, connecting two or more ideas or pieces of
information and recognizing their relationship, and ({\em iii}) interpret and
integrate
ideas and information, focusing on local or global meanings. The Grammar section
is made of 10 items, which measure the ability
of understanding the morphological
and syntactic structure of sentences within a text.

The INVALSI Mathematics Test consisted of 27 items covering four main content
domains: Numbers, Shapes and Figures, Algebra, and Data and Previsions
\citep{INV:092}. The Number content domain consists of understanding (and
operation with) whole numbers, fractions and decimals, proportions, and percentage
values. The Algebra domain requires students the ability to understand, among
others, patterns, expressions and first order equations, and to represent them
through words, tables and graphs. Shapes and Figures covers topics such as
geometric shapes, measurement, location and movement. It entails the ability to
understand coordinate representations, to use spatial visualization skills in
order to move between two and three dimensional shapes, draw symmetrical figures,
and understand and being able to describe rotations, translations, and reflections
in mathematical terms. The Data and Previsions domain includes three main topic
areas: data organization and representation (e.g., read, organize and display data
using tables and graphs), data interpretation (e.g., identify, calculate and
compare characteristics of datasets, including mean, median, mode), and chance
(e.g., judge the chance of an outcome, use data to estimate the chances of future
outcomes).

All items included in the Italian Test are
of multiple choice type, with one
correct answer and three distractors, and are dichotomously scored
(assigning 1 point to  correct answers and 0 otherwise).
The Mathematics Test is also made
of multiple choice items, but it also contains two open questions for which
a partial score of 1 was
assigned to partially correct answers and a score of 2 was given to correct
answers\footnote{For the purposes of the analyses described in the following
sections, the open questions of the Mathematics Test were
dichotomously re-scored ,
giving 0 point to incorrect and partially correct answers and 1
point otherwise.}.\newpage

The two Tests were administered in June 2009, at the end of the
pupils' compulsory educational period. Afterwards, a nationally
representative sample made of 27,592 students was drawn through a
stratified random sampling \citep{INV:093}. From each of the 21
strata (the 21 Italian geographic regions) a sample of schools was
drawn independently and allocation of sample units within each
stratum was chosen proportional to an indicator based on the
standard deviations of certain variables and the stratum sizes
\citep{ney:34}. Classes within schools were then sampled through a
random procedure, with one class sampled in each school, without
taking into account the class size (only schools with less than 10
students were excluded from the sampling procedure). Overall, 1305
schools (and classes) were sampled. Table~\ref {table1} and
Table~\ref{table2} show the distribution of students per gender and
geographic areas, respectively for the Italian Test and the
Mathematics Test\footnote{Foreign students, students with
disabilities and records with missing values were excluded from the
dataset.}.

\begin{table}[hb]\centering
\vspace*{0.5cm}
\small\begin{tabular}{l|ccccc|c} \hline\hline
Gender &    \multicolumn{5}{c|}{Geographic area}         \\
 & NW   &   NE   &   Centre  &   South   &   Islands  & Total \\
\hline
Females     &   1969    &   2203    &   2099    &   2194    &   2173 & 10638\\
Males   &   1922    &   2155    &   2242    &   2258    &   2182    &   10759\\
\hline
Total   &   3891    &   4358    &   4341    &   4452    &   4355    &   21397\\
\hline\hline
\end{tabular}
\caption{\em Distribution of students per gender and geographic area for the
INVALSI Italian Test.}
\label{table1}
\end{table}

\begin{table}[ht!]\centering
\small\begin{tabular}{l|ccccc|c} \hline\hline
Gender &    \multicolumn{5
}{c|}{Geographic area}         \\
 & NW   &   NE   &   Centre  &   South   &   Islands  & Total \\
\hline
Females     &   1606    &   1940    &   1786    &   1884    &   1831    &   8825    \\
Males   &   1538    &   1804    &   1840    &   1866    &   1777    &   9047    \\
\hline
Total   &   3144    &   3744    &   3626    &   3750    &   3608    &   17872   \\
\hline\hline
\end{tabular}
\caption{\em Distribution of students per gender and geographic area
for the INVALSI Mathematics Test.}
\label{table2}\vspace*{0.5cm}
\end{table}
\newpage

Preliminary analyses (see Table~\ref{table3} and Table~\ref{table4})
confirm that students' performances on Test items were different on
account of students' gender and geographic area. Overall, females
performed better than males in the Italian Test, but worse than
males in the Mathematics Test. In both Tests, average percentage
scores per geographic area revealed very diverse levels of
attainment. Generally, students from the Center of Italy performed
better than the rest of the students in the Italian Test.

\begin{table}[!h]\centering
\vspace*{0.5cm}
\small\begin{tabular}{l|ccccc|c} \hline\hline
Gender &    \multicolumn{5
}{c|}{Geographic area}         \\
 & NW   &   NE   &   Centre  &   South   &   Islands  & Overall \\
\hline
Females     &   75.0    &   73.9    &   76.2    &   75.2    &   73.6    &   74.8    \\
Males   &   73.0    &   71.4    &   73.1    &   73.4    &   71.0    &   72.4    \\
\hline
Overall &   74.0    &   72.6    &   74.6    &   74.3    &   72.3    &   73.6    \\
\hline\hline
\end{tabular}
\caption{\em Average percentage score per gender and geographic area for the
INVALSI Italian Test.}
\label{table3}
\end{table}

\begin{table}[ht!]\centering
\small\begin{tabular}{l|ccccc|c} \hline\hline
Gender &    \multicolumn{5
}{c|}{Geographic area}         \\
 & NW   &   NE   &   Centre  &   South   &   Islands  & Overall \\
\hline
Females     &   73.3    &   71.9    &   75.6    &   77.5    &   76.3    &   75.0    \\
Males   &   75.6    &   74.9    &   76.8    &   77.8    &   76.8    &   76.4    \\
\hline
Overall &   74.4    &   73.4    &   76.2    &   77.6    &   76.5    &   75.7    \\
\hline\hline
\end{tabular}
\caption{\em Average percentage score per gender and geographic area for the
INVALSI Mathematics Test.}
\label{table4}
\vspace*{0.5cm}
\end{table}
\section{Methodological approach}\label{sec:methods}
In this section, we illustrate  the methodological approach adopted
to investigate the presence of DIF and the dimension of the latent
structure behind the analysed data. Firstly, we review the basic
model proposed by  \cite{bart:07} and then we extend it to include
DIF effects.

\subsection{Preliminaries}\label{sec:preliminaries}

The multidimensional latent class (LC) IRT models developed by
\cite{bart:07} presents two main differences with respect to the
classic IRT models: (\textit{i}) the latent structure is
multidimensional and (\textit{ii}) it is based on latent variables
that have a discrete distribution. We consider in particular the
version of these models based on the two-parameter (2PL) logistic
parameterisation of the conditional response probabilities.

Let $n$ denote the number of subjects in the sample and suppose that
these subjects answer $r$ dichotomous test items which measure $s$
different latent traits or dimensions. Also let $\mathcal{J}_d$, $d
= 1,\ldots,s$, be the subset of $\mathcal{J}=\{1,\ldots,r\}$
containing the indices of the items measuring the latent trait of
type $d$ and let $r_d$ denoting the cardinality of this subset, so
that $r=\sum_{d=1}^rs_d$. Since we assume that each item measures
only one latent trait, the subsets $\mathcal{J}_d$ are disjoint;
obviously, these latent traits may be correlated. Moreover, adopting
a 2PL parameterisation \citep{bir:68}, it is assumed that

\begin{equation}
\textrm{logit}[p(Y_{ij}=1\mid \bTheta_i = \btheta)] =
\gamma_j\left(\sum_{d=1}^{D} \delta_{jd} \theta_d - \beta_j\right),
\quad i=1,\ldots,n,\:j = 1,\ldots,r. \label{eq:multid2PL}
\end{equation}
In the above expression, $Y_{ij}$ is the random variable
corresponding to the response to item $j$ provided by subject $i$
($Y_{ij}=0,1$ for wrong or right response, respectively). Moreover,
$\beta_j$ and $\gamma_j$ are, respectively, the difficulty and the
discrimination of item $j$, $\bTheta_i = (\Theta_{i1},\ldots,
\Theta_{is})\tr$ is the vector of latent variables corresponding to
the different traits measured by the test items,
$\b\theta=(\theta_1,\ldots,\theta_s)\tr$ denotes one of the possible
realizations of $\b\Theta_i$, and $\delta_{jd}$ is a dummy variable
equal to $1$ if item $j$ belongs to $\mathcal{J}_d$ (and then it
measures the $d$th latent trait) and to 0 otherwise. Finally, a
crucial assumption is that each random vector $\bTheta_i$ has a
discrete distribution with support $\{\b\xi_1,\ldots,\b\xi_k\}$,
which correspond to $k$ latent classes in the population. The
elements of each vector $\b\xi_c$ are denoted by $\xi_{cd}$,
$d=1,\ldots,s$, with $\xi_{cd}$ denoting the ability level of
subjects in latent class $c$ with respect to dimension $d$. Note
that, when $\ga_j=1$ for all $j$, then the above 2PL
parameterisation reduces to a multidimensional Rasch
parameterisation \citep{rasch:61}. At the same time, when the
elements of each support vector $\b\xi_c$ are obtained by the same
linear transformation of the first element, the model is indeed
unidimensional even when $s>1$. The last consideration will be
useful in order to compute $p$-values for the test of
unidimensionality.

As for the conventional LC model \citep{laza:henr:68,good:74}, the
assumption that the latent variables have a discrete distribution
implies the following {\em manifest distribution} of the full
response vector $\b Y_i= (Y_{i1},\ldots,Y_{ir})\tr$:
\begin{equation}
p_i(\b y)=p(\b Y_i=\b y) = \sum_{c=1}^k p_i(\b y\mid
c) \pi_c,\label{equ:1}
\end{equation}
where $\b y=(y_1,\ldots,y_r)\tr$ denotes a realisation of $\b Y_i$,
$\pi_c = p(\bTheta_i=\b\xi_c)$ is the weight of the $c$th latent
class, and
\begin{equation}
p_i(\b y\mid c)=
p(\b Y_i=\b y\mid\b\Theta_i=\b\xi_c)=
\prod_{j=1}^r p(Y_{ij}=y_j \mid
\b\Theta_i=\b\xi_c), \quad c=1,\ldots,k.
\label{eq:cond_prob}
\end{equation}

The specification of the multidimensional LC 2PL model, based on the
assumptions illustrated above, univocally depends on: ({\em i}) the
number of latent classes ($k$), ({\em ii}) the number of the
dimensions ($s$), and ({\em iii}) the way items are associated to
the different dimensions. The last feature is related to the
definition of the subsets $\mathcal{J}_d$, $d = 1,\ldots,s$.
\subsection{Extension for Differential Item Functioning}
DIF occurs when subjects belonging to different groups (commonly
defined by gender, ethnicity, or geographic area) with the same
latent trait level have a different probability of providing a
certain answer to a given item \citep{thi:Ste:wai:93,
clau:et:al:1998, swam:roger:1990}.

Even in the presence of a 2PL parameterisation, it reasonable to suppose that
the main reason of DIF is due to the item difficulty level, which may depend
on the individual characteristics of the respondent. More precisely,
the presence of DIF in the difficulty level of item $j$ may be represented
by shifted values of $\beta_j$ for one group of subjects with respect to another.

Let $z_{gi}$ be a dummy variable which assumes value 1 if subject
$i$ belongs to group $g$ (e.g., that of females) and value 0
otherwise. The number of groups is denoted by $h$, so that, in the
previous expression, $g=1,\ldots,h$. When $s=1$, the 2PL
parameterisation may be extended for DIF by assuming:
\begin{equation}\label{eq:2PLdif}
\hspace*{-0.3cm}
\logit[p(Y_{ij}=1\mid \Theta_i = \theta)] = \gamma_j
\left[\theta - \left(\beta_j+\sum_{g=1}^h\phi_{gj}z_{gi}\right)\right],
\quad i=1,\ldots,n,\:j=1,\ldots,k,
\end{equation}
where $\phi_{gj}$ measures the shift for item $j$ in terms of difficulty.
Therefore, if two subjects have the same ability
level $\th$, but belong to two different groups, say $g_1$ and $g_2$,
the difference between the corresponding conditional probabilities of
a correct response is $\phi_{g_1j}-\phi_{g_2j}$ on the logit scale.
It can be observed that this
difference between logits does not depend on the common latent
trait value $\theta$. In this case, the so-called {\em uniform DIF} arises;
see \cite{thi:Ste:wai:93}, \cite{clau:et:al:1998}, and
\cite{swam:roger:1990}.

Obviously, DIF in the difficulty level may be also introduced in the
multidimensional case and when subjects are classified according to
more criteria, to give an additive structure to the corresponding
DIF effects. More precisely, suppose that, as in our applications,
subjects are grouped according to two criteria and that the first
criterion gives rise to $h_1$ groups, whereas the second gives rise
to $h_2$ groups. Then, as an extension of (\ref{eq:multid2PL}), we
have
\begin{equation} \label{eq:multi2PLdif2}
\textrm{logit}[p(Y_{ij}=1\mid \bTheta_i = \btheta)] =
\gamma_j\left[\sum_{d=1}^{s}
\delta_{jd} \theta_d - \left(\beta_j+
\sum_{g=1}^{h_1} \phi_{gj}^{(1)}z_{hi}^{(1)}+
\sum_{g=1}^{h_2} \phi_{gj}^{(2)}z_{hi}^{(2)}\right)\right],
\end{equation}
for $i=1,\ldots,n$ and $j=1,\ldots,r$. In the above expression, each
dummy variable $z_{gi}^{(1)}$ is equal to 1 if subject $i$
belongs to group $g$ (when the classification of subjects is based on the
first criterion) and to 0 otherwise; $\phi_{gj}^{(1)}$ is the corresponding
DIF parameter. The dummy variables $z_{gi}^{(2)}$ and the parameters
$\phi_{gj}^{(2)}$ are defined accordingly. These parameters may be simply
interpreted as clarified above.\newpage

In the approach proposed in this paper, we rely on assumption
(\ref{eq:multi2PLdif2}) to extend the class of model of
\cite{bart:07} for DIF, even under the 2PL parameterisation and in
the presence of multidimensionality.
\section{Likelihood based inference}\label{sec:ML}
In this section, we deal with the maximum likelihood of the extended
model based on assumption (\ref{eq:multi2PLdif2}) and
with the problem of selecting the number of latent states, and testing hypotheses on
the parameter. The hypotheses of greatest interest in our context are those of
absence of DIF and unimensionality. We also briefly outline the
algorithm for clustering items in unidimensional groups.
\subsection{Maximum likelihood estimation}
Let $\b y_i$, $i=1,\ldots,n$, denote the response configuration
provided by subject $i$. For a given $k$, the parameters of the
proposed model may be estimated by maximizing the log-likelihood
\begin{equation}\label{eq:loglik}
\ell(\boeta) = \sum_i\log[p_i(\b y_i)],
\end{equation}
where $\boeta$ is the vector containing all the free parameters,
and $p_i(\b y)$ is the manifest mass probability function
of $\b y$ defined in (\ref{equ:1}) on the basis of the model parameters.
When subjects are classified according
to only one criterion, an equivalent expression for the log-likelihood
is the following
\begin{equation}\label{eq:loglik_freq}
\ell(\boeta) = \sum_{g=1}^h\sum_{\bl y}n(g,\b y)\log[p_g^*(\b y)],
\end{equation}
where $n(g,\b y)$ is the frequency, in the sample, of subjects who
belong to group $g$ and provide response configuration $\b y$, and
$p_g^*(\b y)$ is the manifest probability of $\b y$ for the subjects.
Moreover, the sum $\sum_{\bl y}$ is extended to all response configurations
observed at least once. Similar expressions result when subjects are classified
according to more criteria.

About the vector $\b\eta$, we clarify that it contains the item parameters
$\be_j$ (difficulty) and $\ga_j$ (discriminating index), and $\phi_{gj}$
(DIF parameters), the parameters $\xi_{cd}$ (ability levels) and $\pi_c$
(corresponding weights). However, to make  the model identifiable, we adopt
the constraints
\[
\beta_{j_d}=0,\:\gamma_{j_d}=1,\quad d=1,\ldots,s,
\]
with $j_d$ denoting a reference item for the $d$-th dimension
(usually, but not necessarily, the first one in the group). When
subjects are classified according to only one criterion, we have
\begin{equation}
\phi_{1j}=0,\quad j=1,\ldots,r,\label{eq:ident_cons}
\end{equation}
where the first group is taken as reference group. In this way, for
each item $j$, with $j\in(\cg J_d\setminus\{j_d\})$, the parameter
$\be_j$ is interpreted in terms of differential difficulty level of
this item with respect to item $j_d$; similarly, $\ga_j$, is
interpreted in terms of ratio between the discriminant index of item
$j$ and that of item $j_d$. Finally, for $g>1$, $\phi_{gj}$
corresponds to the differential difficultly level of group $g$, with
respect to the first group, for item $j$. When subjects are
classified according to, say, two criteria and assumption
(\ref{eq:multi2PLdif2}) is adopted, then the identifiability
constraints
\[
\phi^{(1)}_{1j}=\phi^{(2)}_{1j}=0,\quad j=1,\ldots,r,
\]
must be used instead in (\ref{eq:ident_cons}).

Considering the above identifiability constraints, when $k>2$ and subjects are classified
according to a single criterion with regard to DIF, the number of free parameters
collected in $\boeta$ is equal to
$$
\#{\rm par}=(k-1) + ks + 2(r-s) + r(h - 1),
$$
since there are $k-1$ free latent class probabilities, $ks$ free
ability parameters $\xi_{cd}$, $r-s$ free difficulty parameters and
discriminant indices, and $r(h-1)$ free DIF parameters. For $k=1,2$,
the proposed model does not pose any restriction over the LC model
and then we have $\#{\rm par}=(k-1) + kr + r(h-1)$. The number of
parameters is simply modified when subjects are classified according
to more criteria. For instance, when the classification is based on
two criteria, and then assumption (\ref{eq:multi2PLdif2}) holds, the
term $r(h - 1)$ in $\#{\rm par}$ need to be substituted by
$r(h_1+h_2-2)$.

In order to maximise the log-likelihood $\ell(\boeta)$, we make use of the
Expectation-Maximization (EM) algorithm \citep{demp:lair:rubi:77},
which is
implemented along the same lines as in \cite{bart:07}. This algorithm
is briefly described in Appendix 1; a {\sc Matlab} implementation is available
from the authors upon request. The maximum likelihood estimate of $\b\eta$,
obtained from maximisation of $\ell(\boeta)$, is denoted by $\hat{\boeta}$.

After the parameter estimation, each subject $i$ can be allocated to one of
the $k$ latent classes on the basis of the response pattern $\b y_i$ he/she
provided. The most common approach is to assign the subject to the
class with the highest posterior probability. On the basis of the parameter
estimates, the posterior probability is computed as
\begin{equation}
\hat{q}_i(c\mid\b y_i) = \hat{p}_i(\bTheta_i=\b\xi_c
\mid\b Y_i=\b y_i) =
\frac{\hat{p}_i(\b y_i|\b\xi_c)\hat{\pi}_c}
{\sum_{h=1}^k\hat{p}_i(\b y_i|\b\xi_h)\hat{\pi}_h}, \quad c=1,\ldots,C.
\end{equation}
\subsection{Choice of the number of latent classes, hypothesis testing,
and dimensionality assessment}\label{sec:unidim}
In analysing a dataset by the model described in Section
\ref{sec:methods}, a crucial point is the choice of the number of
latent classes $k$. To this aim, we rely on the Bayesian Information
Criterion (BIC) of \cite{sch:78}. On the basis of this criterion,
the selected number of classes is the one corresponding to the
minimum value of
\[
BIC = -2\ell(\hat{\b\eta})+\log(n)\#{\rm par}.
\]
In practice, we fit the model for increasing values of $k$ until
$BIC$ does not start to increase and then we take the previous value
of $k$ as the optimal one.

Once the number of latent states has been selected, it is of
interest to test several hypotheses on the parameters. To this aim,
we can follow the general likelihood ratio (LR) approach. For a
hypothesis of type $H_0:\b f(\b\eta)=\b 0$, where $\b 0$ denotes a
column vector of zeros of suitable dimension, this approach is based
on the statistic
\begin{equation}
D = -2[\ell(\hat{\b\eta}_0)-\ell(\hat{\b\eta})],\label{eq:LR}
\end{equation}
which, under the usual regularity conditions, has null asymptotic distribution
of $\chi^2_m$ type, where $m$ is the number of constraints imposed by $H_0$.
An alternative approach is based on the
Wald test which is based on the statistic
\begin{equation}
W=\b f(\hat{\b\eta})\tr\b G(\hat{\b\eta})\b f(\hat{\b\eta}),\label{eq:Wald}
\end{equation}
where $\b G(\b\eta)$ is a suitable matrix computed on the basis of
the Jacobian of $\b f(\b\eta)$ and the information matrix of the
model. It is well known that the two approaches are asymptotically
equivalent, and that, differently from the LR approach, the one
based on the statistic $W$ only requires to fit the larger model,
but also to compute the information matrix of the model, which may
be rather complex.

On the basis of the above approach, we can test the hypothesis of absence of
DIF. In this case, the null hypothesis is
\[
H_0:\phi_{gj}=0,\quad g=2,\ldots,h,\:j=1,\ldots,r,
\]
or
\begin{equation}
H_0:\phi_{2j}^{(1)}=\cdots=\phi_{h_1j}^{(1)}=
\phi^{(2)}_{2j}=\cdots=\phi^{(2)}_{h_2j}=0,\quad j=1,\ldots,r,
\label{eq:noDIF}
\end{equation}
when subjects are classified according to two criteria for what concerns
DIF. Then, to test $H_0$, we have to fit the model with and without DIF and compare
the corresponding log-likelihoods by (\ref{eq:LR}). Thus, if the obtained
value of test statistic is higher than a suitable percentile of the $\chi^2_m$
distribution, with $m=r(h-1)$, we reject $H_0$ and can state that
there is evidence of DIF.

The above approach may also be used for the hypothesis that a group of items
measure only one latent trait, that is unidimensionality, against the hypothesis
that the same group is multidimensional. In the case of two dimensions,
for instance,
we have to compare by the LR statistic (\ref{eq:LR}) the model in which these
dimensions are kept distinct with the model
in which these dimensions are collapsed.
Under the null hypothesis of unidimensionality, this test statistic has
an asymptotic
distribution of $\chi^2_m$ type, with $m=k-2$. This is because, as mentioned in
Section \ref{sec:preliminaries}, unidimensionality holds when the ability level
for the second dimension may be obtained by the same linear transformation of the
ability level for the first dimension, for every latent class $c$. Obviously,
this test makes only sense when $k>2$ and, in general, may also be performed by
a Wald statistic of type (\ref{eq:Wald}), once the function $\b f(\b\eta)$ has been
suitably defined; see \cite{bart:07} for details.

By repeating the test for unimensionality mentioned above in a
suitable way, we can cluster items so that items in the same group
measure the same ability. On the basis of this principle,
\cite{bart:07} proposed a hierarchical clustering algorithm that we
also apply for the extended models here proposed, which account for
DIF. This algorithm builds a sequence of nested models: the most
general one is that with a different dimension for each item
(corresponding to the classic LC model in absence of DIF) and the
most restrictive model is that with only one dimension common to all
items (unidimensional model). The clustering procedure performs
$s-1$ steps. At each step, the Wald test statistic for
unidimensionality is computed for every pair of possible
aggregations of items (or groups of items). The aggregation with the
minimum value of the statistic (or equivalently the highest
$p$-value) is then adopted and the corresponding model fitted before
going to the next step. A similar strategy could be based on the LR
statistic, but in this case we would be required to fit a much
higher number of models. A {\sc Matlab} implementation of this
algorithm is also available from the authors upon request.

The output of the above clustering algorithm may be displayed
through a dendrogram that shows the deviance between the initial
($k$-dimensional) LC model and the model selected at each step of
the clustering procedure. Obviously, the results of a cluster
analysis based on a hierarchical procedure depend on the adopted
rule to cut the dendrogram, which may be chosen according to several
criteria. A rule that may be adopted to cut the dendrogram is based
on the increase of a suitable information criterion, such as BIC,
with respect to the initial or the previous fitted model. A negative
increase of BIC means that the new model reaches a better compromise
between goodness-of-fit and parsimony than the model used as a
comparison term (i.e., the initial or the previous one). The
dendrogram is cut when the item aggregation does not give any
additional advantages, that is, in correspondence with the last step
showing a negative increase.
\section{Application to the INVALSI dataset}\label{sec:appl}
In this section, we apply the extended class of models to the data
collected by the two INVALSI Tests. For the purposes of the
analysis, the 30 items which assess reading comprehension within the
Italian Test are kept distinct from the 10 items which assess
grammar competency, as the two sections deal with two different
competencies. Besides, since we do not have any prior information on
item discrimination power, we choose the 2PL parameterisation and,
regarding the way of taking DIF effects into account, we consider
subjects classified according to gender ($h_1=2$ categories: Males,
Female) and geographical area ($h_2=5$ categories: NorthWest,
NorthEast, Centre, South, Islands). Then, the adopted
parameterisation is the same as in (\ref{eq:multi2PLdif2}). The
categories Males and NorthWest are taken as reference categories.

In the following, we deal with the selection of the number of latent
classes, with the problem of testing the hypothesis of absence of
DIF, and with the issue of clustering items.

\subsection{Selection of the number of classes}\label{sec:selection}
In order to choose the number of latent classes we proceed as
described in Section \ref{sec:unidim} and fit the model in the
multidimensional version, in which each item is assumed to measure a
single ability, for values of $k$ from 1 to 9. The maximum value of
$k$ is chosen to be equal to 9 as it is the first value for which
$BIC$ is higher than that associated to the previous value of $k$
for all Test sections. The results of this preliminary fitting are
reported in Table \ref{table5}.

\begin{table}[!ht]\centering
\vspace*{0.5cm}
{\small
\begin{tabular}{l|rcr|rcr|rcr}
\hline\hline
$k$   &     \multicolumn3c{Reading comprehension}      &
\multicolumn3c{Grammar}    &     \multicolumn3c{Mathematics}   \\
\hline
           &       \multicolumn1c{$\ell(\hat{\b\eta})$}   &
           \multicolumn1c{$\#{\rm par}$}       &      \multicolumn1c{$BIC$}  &
           \multicolumn1c{$\ell(\hat{\b\eta})$}    &
           \multicolumn1c{$\#{\rm par}$}     &      \multicolumn1c{$BIC$}   &
\multicolumn1c{$\ell(\hat{\b\eta})$}    &       \multicolumn1c{$\#{\rm par}$}
    &      \multicolumn1c{$BIC$}    \\
\hline
1   &   -350,474    &   180 &   702,743 &   -100,842    &   60  &   202,282
&   -242,111    &   162 &    485,808    \\
2   &   -329,109    &   211 &   660,323 &   -95,580 &   71  &   192,899
&   -224,506    &   190 &    450,873    \\
3   &   -326,171    &   242 &   654,760 &   -95,645 &   82  &   192,110
&   -221,976    &   218 &    446,090    \\
4   &   -325,516    &   273 &   653,750 &   -95,580 &   93  &   192,090
&   -220,936    &   246 &    444,280    \\
5   &   -324,970    &   304 &   \bf 652,970 &   -95,517 &   104 &   \bf 192,070
&   -220,032    &   274 &    442,750    \\
6   &   -324,863    &   335 &   653,070 &   -95,470 &   115 &   192,090
&   -219,619    &   302 &    442,190    \\
7   &   -324,764    &   366 &   653,178 &   -95,464 &   126 &   192,184
&   -219,248    &   330 &    441,730    \\
8   &   -324,684    &   397 &   653,327 &   -95,454 &   137 &   192,274
&   -218,977    &   358 &    \bf 441,460    \\
9   &   -324,583    &   428 &   653,436 &   -95,429 &   148 &   192,334
&   -218,846    &   386 &    441,470    \\
\hline\hline
\end{tabular}}
\caption{\em Log-likelihood, number of parameters and BIC values for $k = 1,
\ldots, 9$ latent classes for the Reading Comprehension and the Grammar sections
of the Italian Test and for the Mathematics Test; in boldface is the smallest BIC
value for each type of Test.}
\label{table5}\vspace*{0.5cm}
\end{table}

On the basis of BIC, we choose $k = 5$ classes both for the Reading
Comprehension and the Grammar sections of the Italian Test. As regards to
Mathematics Test, despite $k=8$ being the optimal number of classes, we choose
$k=3$, as for each number of classes greater than 3 the model becomes almost
nonidentifiable, in the sense that the corresponding information matrix is
close to be singular.
We recall that this matrix is crucial for performing the Wald test for
unidimensionality.
\subsection{Testing absence of DIF}
As previously specified, we define two groups of students on the basis of
gender and
geographic area. The null hypothesis of no (uniform) DIF is formulated
as in (\ref{eq:noDIF}).
At this regard, Table~\ref{table6} shows the LR statistic, computed as in
(\ref{eq:LR}), between the 2PL multidimensional model with uniform DIF based
on assumption (\ref{eq:multi2PLdif2}) and the 2PL multidimensional model
based on assumption (\ref{eq:multid2PL}).

\begin{table}[ht!]\centering\vspace*{0.5cm}
\small\begin{tabular}{l|ccc}
\hline\hline
    &   Deviance    &   $p$-value   \\
\hline
Reading Compr.  &   1579.702 &  $<$0.001    \\
Grammar &   1313.427  & $<$0.001    \\
Mathematics &   2183.573 &  $<$0.001    \\
\hline\hline
\end{tabular}
\caption{\em Deviance of the multidimensional 2PL model with uniform DIF with respect to the multidimensional 2PL model with no DIF for the Italian Test - Reading Comprehension section and Grammar section - and the Mathematics Test.}
\label{table6}\vspace*{0.5cm}
\end{table}

According to these results, the assumption of no
DIF is strongly
rejected for both sections of the Italian Test and for the Mathematics Test.
Therefore, in Table~\ref{table7}, Table~\ref{table8}, and Table~\ref{table9}
we provide the estimates of the DIF coefficients ($\phi_{gj}^{(1)}$ and
$\phi_{gj}^{(2)}$).
We recall that each of these coefficients represents the difference, in terms
of difficulty of an item, between one group of subjects with respect to the
reference group, given the same ability level.

\begin{table}[!ht]\centering
\vspace*{0.5cm}
{\small
\begin{tabular}{p{0.8cm} p{1.6cm}p{1.6cm}p{1.6cm}p{1.6cm}p{1.6cm}}
\hline\hline
Item    &   Females &  NorthEast    & Centre    &  South    & Islands \\
\hline
R1  &   -0.018          &   -0.051          &   -0.262$^{***}$  &   -0.173$^{**}$   &    $\:$0.032          \\
R2  &   -0.322$^{***}$  &   $\:$0.132           &   -0.005          &   -0.073          &    $\:$0.170          \\
R3  &   $\:$0.021           &   $\:$0.211$^{*}$    &   -0.057          &   $\:$0.212$^{*}$            &  $\:$0.131           \\
R4  &   -0.447$^{***}$          &   $\:$0.253$^{*}$       &   $\:$0.291$^{**}$           &   $\:$0.428$^{***}$            &  $\:$0.613$^{***}$           \\
R5  &   -0.377$^{***}$      &   $\:$0.192$^{*}$           &   $\:$0.094           &   $\:$0.110            &  $\:$0.227$^{**}$           \\
R6  &   $\:$0.117$^{**}$           &   $\:$0.132           &   $\:$0.153$^{*}$           &   $\:$0.305$^{***}$            &  $\:$0.457$^{***}$           \\
R7  &   -0.332$^{***}$  &   $\:$0.083           &   $\:$0.040           &   $\:$0.196$^{**}$            &  $\:$0.433$^{***}$           \\
R8  &   -0.072$^{*}$    &   $\:$0.002           &   $\:$0.127$^{*}$          &   $\:$0.229$^{***}$            &  $\:$0.159$^{**}$           \\
R9  &   -0.170$^{***}$  &   $\:$0.046           &   $\:$0.008           &   $\:$0.078            &  $\:$0.141$^{**}$           \\
R10 &   -0.340$^{***}$  &   $\:$0.320$^{*}$           &   -0.291$^{*}$    &   -0.420$^{**}$   &    -0.581$^{***}$ \\
R11 &   -0.159$^{***}$  &   $\:$0.038           &   $\:$0.075           &   $\:$0.279$^{***}$            &  $\:$0.415$^{***}$       \\
R12 &   -0.148$^{***}$ &    $\:$0.069           &   -0.038          &   $\:$0.277$^{***}$           &    $\:$0.227$^{***}$          \\
R13 &   -0.057$^{*}$    &   $\:$0.003           &   $\:$0.003           &   -0.035          &    $\:$0.111$^{*}$          \\
R14 &   $\:$0.096           &   $\:$0.019           &   -0.060          &   $\:$0.176$^{*}$            &  $\:$0.159$^{*}$           \\
R15 &   $\:$0.001           &   -0.026          &   -0.079          &   -0.079          &    $\:$0.028          \\
R16 &   -0.352$^{***}$  &   $\:$0.270$^{**}$           &   -0.387$^{***}$  &   -0.682$^{***}$  &    -0.661$^{***}$ \\
R17 &   -0.074$^{**}$   &   $\:$0.058           &   -0.065          &   -0.017          &    $\:$0.067          \\
R18 &   $\:$0.109$^{**}$           &   $\:$0.036           &   $\:$0.075           &   $\:$0.232$^{***}$            &  $\:$0.350$^{**}$       \\
R19 &   $\:$0.260$^{**}$          &   $\:$0.044           &   -0.075          &   -0.480$^{***}$  &    -0.566$^{***}$ \\
R20 &   $\:$0.029           &   $\:$0.049           &   $\:$0.283$^{***}$           &   $\:$0.207$^{**}$            &  $\:$0.236$^{**}$           \\
R21 &   -0.195$^{***}$  &   -0.068          &   $\:$0.018           &   $\:$0.327$^{***}$           &    $\:$0.290$^{**}$          \\
R22 &   -0.193$^{***}$  &   $\:$0.020           &   -0.022          &   $\:$0.216$^{***}$           &    $\:$0.334$^{***}$          \\
R23 &   -0.254$^{***}$  &   $\:$0.050           &   $\:$0.025           &   $\:$0.431$^{***}$            &  $\:$0.441$^{***}$           \\
R24 &   -0.245$^{*}$   &   $\:$0.223           &   -0.282$^{*}$    &   -0.216    &    -0.067         \\
R25 &   -0.068          &   -0.043          &   $\:$0.001           &   -0.173$^{*}$    &    -0.056         \\
R26 &   -0.319$^{***}$  &   $\:$0.053           &   -0.106    &   -0.158$^{**}$   &    $\:$0.160$^{**}$           \\
R27 &   -0.239$^{***}$  &   -0.116          &   -0.079          &   $\:$0.150           &    $\:$0.313$^{***}$          \\
R28 &   -0.286$^{***}$  &   $\:$0.094           &   -0.105          &   -0.215$^{**}$    &    -0.014         \\
R29 &   -0.179$^{***}$  &   -0.071          &   -0.117          &   $\:$0.008           &    $\:$0.252 $^{**}$         \\
R30 &   -0.405$^{***}$  &   $\:$0.026           &   -0.119          &   $\:$0.182$^{*}$           &    $\:$0.309$^{***}$          \\
\hline\hline
\end{tabular}
\caption{\em Estimated DIF coefficients for the Italian Test items  - Reading
Comprehension Section; significance at levels 0.001 (***), 0.01 (**), 0.05 (*).}
\label{table7}}\vspace*{0.5cm}
\end{table}

The results in the previous tables show that the Italian Test
generally favours girls; conversely, the items of the Mathematics
Test tend to favour boys. When taking into account students'
geographic area, we observe that the incidence of items affected by
DIF is, on the whole, stronger for the southern regions (Islands
included) than the central and northeastern regions, with a higher
proportion of items significantly affected by DIF when accounting
for the former geographic areas, both in the Italian Test and in the
Mathematics Test. Specifically, as for the two sections of the
Italian Test, the analysis shows that students from the southern
regions tend to have a lower chance to answer the items correctly
than students from the other Italian regions. On the contrary,
Mathematics Test items generally tend to favour students from the
South of Italy.
\subsection{Dimensionality assessment}
\begin{figure}[h!]\centering
\includegraphics [width=15cm] {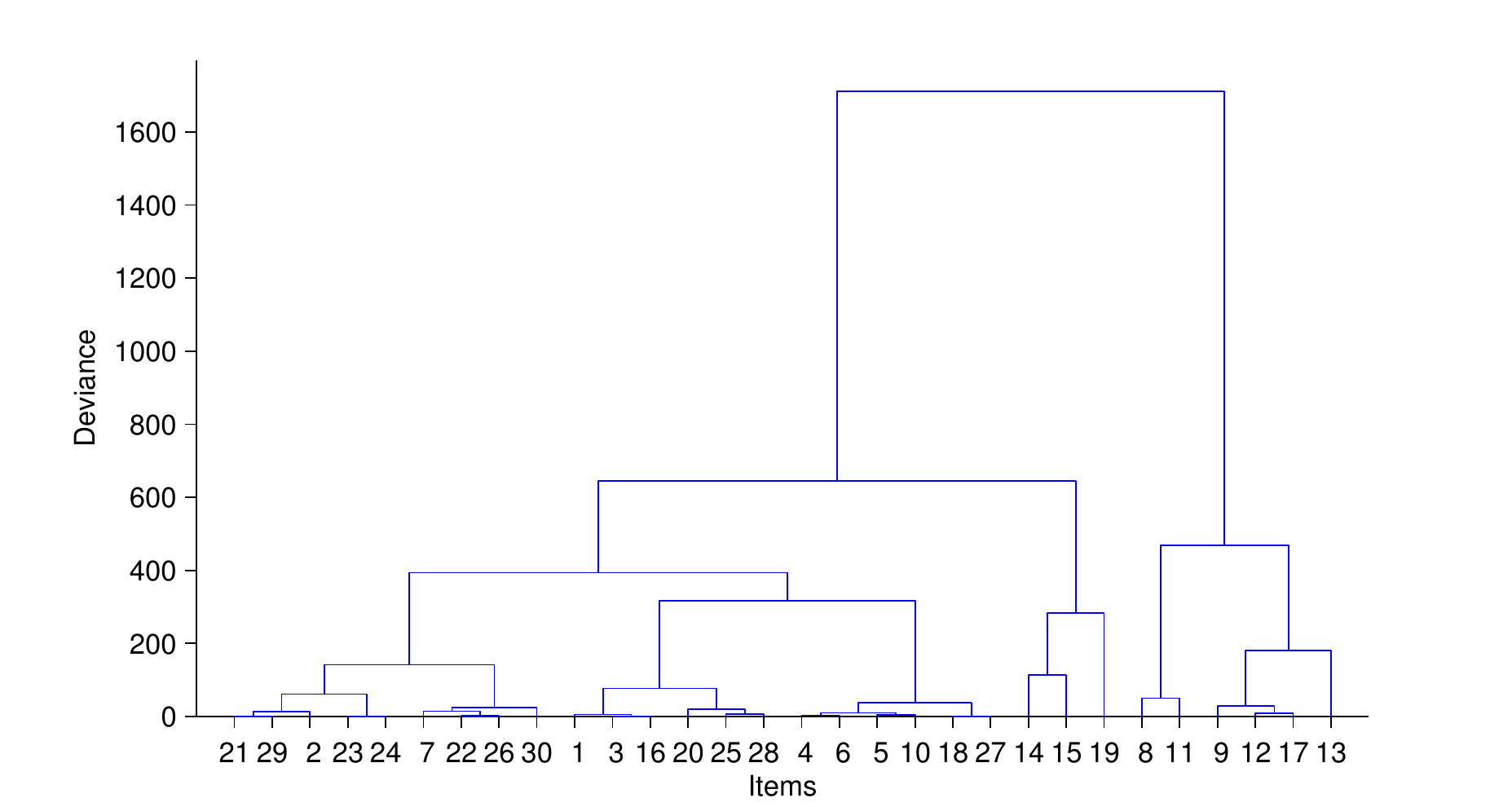}
\caption{\em Dendrogram for the Italian Test - Reading Comprehension Section}\vspace*{0.5cm}
\label{figure1}
\end{figure}

\begin{figure}[!h]\centering
\includegraphics [width=15cm] {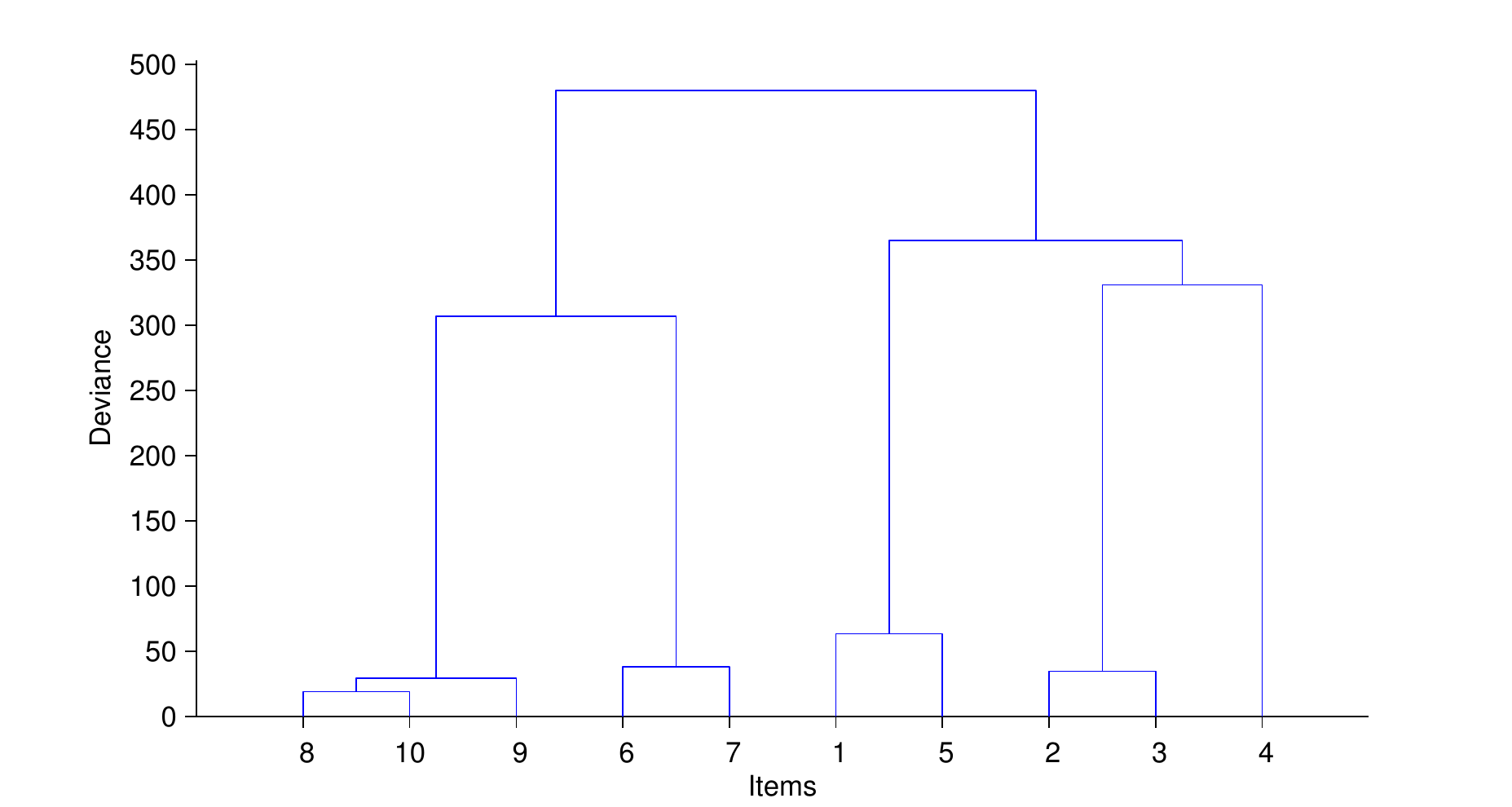}
\caption{\em Dendrogram for the Italian Test - Grammar Section}\vspace*{0.5cm}
\label{figure2}
\end{figure}

\begin{figure}[!h]\centering
\includegraphics [width=15cm] {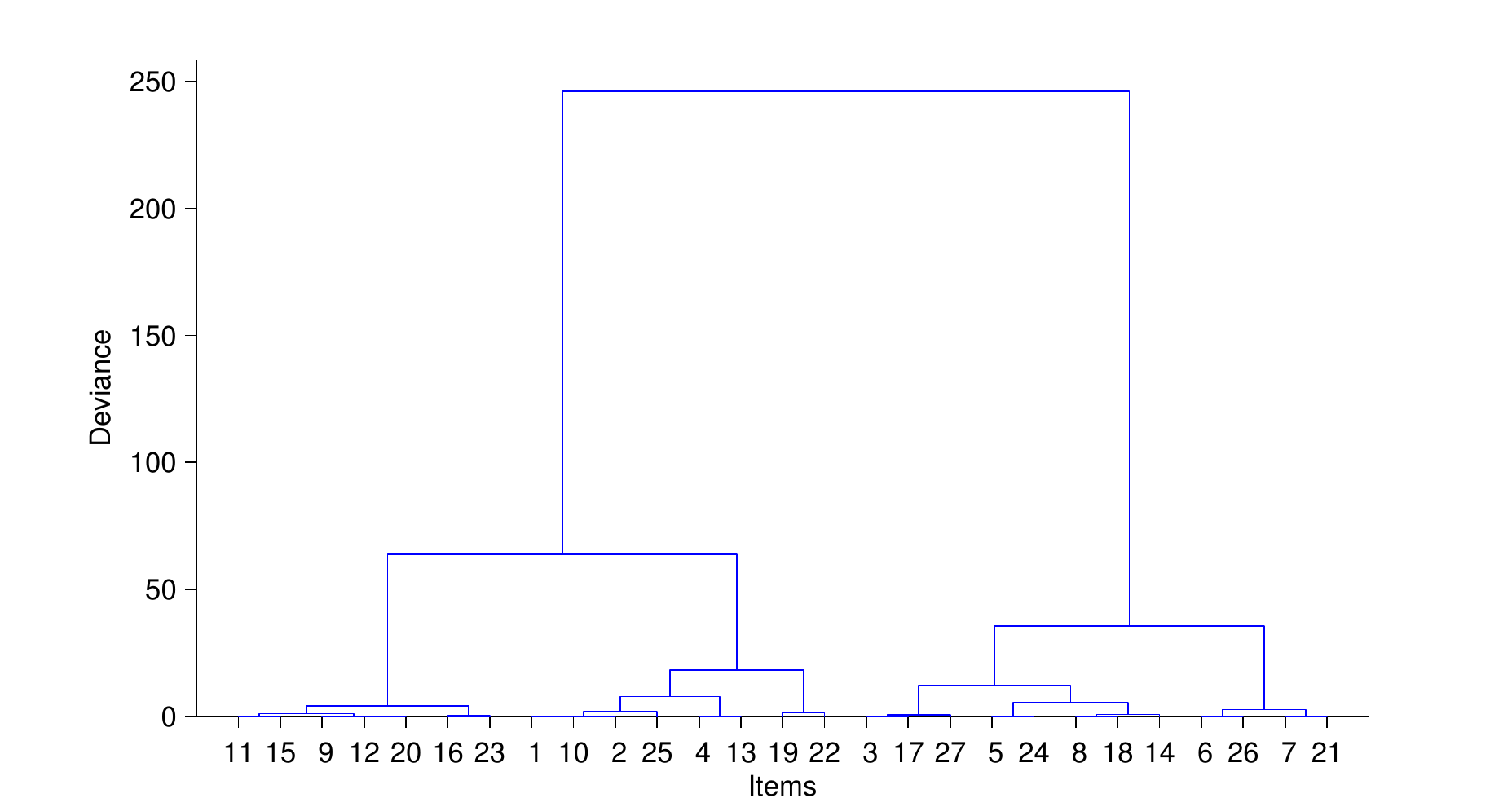}
\caption{\em Dendrogram for the Mathematics Test}\vspace*{0.5cm}
\label{figure3}
\end{figure}

Once the model which includes DIF has been adopted, with a specific
$k$ for each Test section (as defined in Section
\ref{sec:selection}), we performed the item clustering algorithm
described in Section \ref{sec:unidim}. The output of this algorithm
is represented by the dendrograms in Figures~\ref{figure1},
\ref{figure2}, and \ref{figure3}, which are referred, respectively,
to the Reading Comprehension section of the Italian Test, to the
Grammar section of the same Test, and to the Mathematics Test.

\begin{table}[h!]\centering\vspace*{0.5cm}
{\small
\begin{tabular}{p{0.8cm} p{1.6cm}p{1.6cm}p{1.6cm}p{1.6cm}p{1.6cm}}
\hline\hline
Item    &   Females &     NorthEast &  Centre   &   South   & Islands  \\
\hline
G1  &   -0.404$^{***}$  &   $\:$0.133           &   $\:$0.073       &   $\:$0.129           &    $\:$0.428$^{***}$          \\
G2  &   -0.272$^{***}$  &   $\:$0.156$^{*}$           &   $\:$0.060           &   $\:$0.051            &  $\:$0.114           \\
G3  &   -0.137$^{***}$  &   $\:$0.059           &   -0.191$^{**}$   &   -0.198$^{**}$   &    -0.002     \\
G4  &   -0.052          &   $\:$0.323$^{***}$           &   $\:$0.004           &   -0.679$^{***}$  &    -0.350$^{***}$ \\
G5  &   -0.328$^{***}$  &   $\:$0.118$^{*}$           &   -0.111$^{*}$    &   -0.141$^{**}$   &    $\:$0.126$^{*}$           \\
G6  &   -0.261$^{***}$  &   $\:$0.362$^{***}$           &   -0.043          &   -0.312$^{***}$ &     $\:$0.072          \\
G7  &   -0.323$^{***}$  &   $\:$0.197$^{***}$           &   -0.131$^{*}$   &   -0.287$^{***}$  &    -0.011         \\
G8  &   -0.309$^{***}$  &   $\:$0.060           &   $\:$0.144           &   $\:$0.099            &  $\:$0.524$^{***}$          \\
G9  &   -0.205$^{*}$  &   -0.084          &   -0.067          &   $\:$0.290$^{*}$       &    $\:$0.705$^{***}$          \\
G10 &   -0.167$^{*}$    &   $\:$0.269$^{*}$           &   -0.280$^{*}$    &   -0.504$^{***}$  &    -0.324$^{*}$  \\
\hline\hline
\end{tabular}}
\caption{\em Estimated DIF coefficients for the Italian Test items  - Grammar
Section; significance at levels 0.001 (***), 0.01 (**), 0.05 (*).
\label{table8}}\vspace*{0.5cm}
\end{table}

\begin{table}[!h]\centering\vspace*{0.5cm}
{\small
\begin{tabular}{p{0.8cm} p{1.6cm}p{1.6cm}p{1.6cm}p{1.6cm}p{1.6cm}}
\hline\hline
Item    &   Females  &  NorthEast    & Centre   &  South     & Islands  \\
\hline
M1  &   $\:$0.092           &   $\:$0.013           &   -0.193$^{**}$   &   -0.426$^{***}$  &    -0.423$^{***}$ \\
M2  &   -0.027          &   $\:$0.058           &   -0.051    &   -0.050   &    $\:$0.045          \\
M3  &   $\:$0.023           &   $\:$0.058           &   -0.044          &   -0.277$^{***}$  &    -0.200$^{***}$ \\
M4  &   $\:$0.008           &   $\:$0.076$^{*}$           &   -0.021          &   -0.030          &    $\:$0.006          \\
M5  &   $\:$0.024           &   $\:$0.102           &   $\:$0.109           &   $\:$0.057            &  -0.060          \\
M6  &   -0.002          &   $\:$0.012           &   -0.102          &   $\:$0.072           &    $\:$0.076          \\
M7  &   $\:$0.036           &   -0.097          &   $\:$0.073           &   -0.090          &    -0.139$^{***}$   \\
M8  &   $\:$0.022           &   -0.058    &   $\:$0.001           &   $\:$0.144            &  $\:$0.148           \\
M9  &   $\:$0.071$^{***}$           &   -0.022          &   -0.056$^{*}$    &   -0.078$^{**}$   &    -0.027         \\
M10 &   $\:$0.102$^{***}$           &   $\:$0.023           &   -0.041   &   -0.083$^{**}$  &    -0.077$^{**}$ \\
M11 &   $\:$0.029$^{*}$           &   $\:$0.031           &   -0.120$^{***}$  &   -0.310$^{***}$  &    -0.310$^{***}$ \\
M12 &   $\:$0.087$^{***}$            &   $\:$0.024           &   -0.057$^{*}$    &   -0.072  &    $\:$0.022      \\
M13 &   $\:$0.055$^{***}$           &   $\:$0.101$^{***}$           &   -0.027          &   -0.073$^{**}$   &    -0.024         \\
M14 &   $\:$0.052$^{**}$           &   $\:$0.065$^{**}$           &   -0.031          &   -0.166$^{***}$  &    -0.193$^{***}$ \\
M15 &   $\:$0.052$^{**}$           &   $\:$0.037           &   -0.012          &   $\:$0.047            &  -0.002          \\
M16 &   $\:$0.161$^{***}$           &   $\:$0.030           &   $\:$0.019           &   -0.008           &  -0.019          \\
M17 &   $\:$0.164$^{***}$           &   -0.001          &   -0.056   &   -0.060    &    $\:$0.018          \\
M18 &   $\:$0.049$^{***}$           &   -0.001          &   -0.023          &   -0.022          &    -0.025         \\
M19 &   -0.008          &   -0.006          &   $\:$0.032           &   $\:$0.103$^{***}$            &  $\:$0.183$^{***}$           \\
M20 &   -0.024          &   $\:$0.029           &   -0.007          &   -0.055$^{*}$    &    -0.006         \\
M21 &   $\:$0.112$^{***}$           &   $\:$0.104$^{**}$           &   $\:$0.023       &   -0.034          &    -0.022         \\
M22 &   $\:$0.033           &   $\:$0.078$^{*}$           &   -0.036          &   $\:$0.001            &  -0.060$^{*}$    \\
M23 &   $\:$0.013           &   $\:$0.049$^{*}$           &   -0.049$^{*}$   &   -0.062$^{**}$   &    -0.057$^{**}$  \\
M24 &   -0.012          &   -0.008          &   -0.097$^{***}$  &   -0.143$^{***}$ &     -0.125$^{***}$ \\
M25 &   -0.035$^{**}$   &   -0.013          &   $\:$0.033           &   $\:$0.032            &  $\:$0.153$^{***}$           \\
M26 &   $\:$0.087$^{***}$      &   $\:$0.089$^{**}$           &   -0.058   &   -0.163$^{***}$  &    -0.083$^{*}$  \\
M27 &   $\:$0.010           &   $\:$0.093$^{**}$    &   -0.074    &   -0.269$^{***}$  &    -0.312$^{***}$ \\
\hline\hline
\end{tabular}}
\caption{\em Estimated DIF coefficients for the Mathematics Test items;
significance at levels 0.001 (***), 0.01 (**), 0.05 (*).
\label{table9}}\vspace*{0.5cm}
\end{table}

Following what outlined in Section 3.3, we adopt as a criterion to cut the
dendrogram the one based on BIC. In particular, since BIC tends to select more
parsimonious models than other criteria (in particular with large sample sizes),
and for consistency with the criterion applied to select the
number of latent classes, we rely on the increase of BIC with respect to the
initial model (i.e., the model with one dimension for each item).
The values of the increase of BIC with respect to
the initial model are shown in Table \ref{table10}; note that the number
of steps of the clustering algorithm depends on the number of items which
are analysed.\newpage

The results in Table \ref{table10} show that, with the adopted cut
criterion and the chosen number of latent classes, the assumption of
unidimensionality is not reasonable for both sections of the Italian
Test, and in particular for the Grammar section, whereas it is
reasonable for the Mathematics Test. Indeed, there is evidence of $s
= 2$ groups of items in the Reading Comprehension Section of the
Italian Test, $s = 5$ groups of items in the Grammar Section of the
Italian Test, and $s = 1$ group of items in the Mathematics Test.
The 2 groups observed within the Reading Comprehension Section of
the Italian Test are made of 24 and 6 items, corresponding to
different, although correlated, dimensions which may be identified
as the ability to: ({\em i}) make sense of worlds and sentences in
the text and recognize meaning connections among them (24 items) and
({\em ii}) interpret, integrate and make inferences from a written
text (6 items). As regards to the Grammar Section of the Italian
Test, the 5 groups of items correspond to the ability to: ({\em i})
recognize verb forms (1 item), ({\em ii}) recognize the meaning of
connectives within a sentence (3 items), ({\em iii}) recognize
grammatical categories (2 items), ({\em iv}) make a difference
between clauses within a sentence (2 items), and ({\em v}) recognize
the meaning of punctuation marks (2 items).

\begin{table}[!ht]\centering\vspace*{0.5cm}
{\small
\begin{tabular}{l|c|ccccccc}
\hline\hline
$h$ &   $s$ &   \multicolumn3c{Increase BIC with respect to initial model} \\
\hline
    &       &   Reading compr. & Grammar & Mathematics          \\
\hline
1   &   29  &       $\;$-29.8   &   -10.755 &   $\;\;$-9.791    \\
2   &   28  &       $\;$-59.6   &   -30.642 &   $\;$-19.581 \\
3   &   27  &           $\;$-89.0   &   -55.184 &   $\;$-29.371 \\
4   &   26  &       -118.4  &   -81.534 &   $\;$-39.158 \\
5   &   25  &       -147.4  &   -86.135 &   $\;$-48.942 \\
6   &   24  &       -176.4  &   {\bf 127.498} &   $\;$-58.723 \\
7   &   23  &       -205.4  &   121.555 &   $\;$-68.499 \\
8   &   22  &       -234.0  &   125.539 &   $\;$-78.275 \\
9   &   21  &       -262.6  &   210.922 &   $\;$-88.043 \\
10  &   20  &       -291.0  &   --  &   $\;$-97.805 \\
11  &   19  &       -319.1  &   --  &   -107.550    \\
12  &   18  &       -346.7  &   --  &   -117.241    \\
13  &   17  &       -374.0  &   --  &   -126.792    \\
14  &   16  &       -399.4  &   --  &   -136.315    \\
15  &   15  &       -424.5  &   --  &   -145.827    \\
16  &   14  &       -449.3  &   --  &   -155.235    \\
17  &   13  &       -470.0  &   --  &   -164.625    \\
18  &   12  &       -488.7  &   --  &   -173.480    \\
19  &   11  &       -507.3  &   --  &   -181.965    \\
20  &   10  &       -522.0  &   --  &   -190.288    \\
21  &   9   &       -514.0  &   --  &   -197.723    \\
22  &   8   &          -516.5   & --        &   -203.161    \\
23  &   7   &       -508.3  &   --  &   -206.950    \\
24  &   6   &       -435.6  &   --  &   -199.422    \\
25  &   5   &       -430.4  &   --  &   -181.022    \\
26  &   4   &       -384.1  &   --  &   $\;\;\;$-8.600  \\
27  &   3   &       -339.4  &   --  &   --  \\
28  &   2   &       -193.6  &   --  &   --  \\
29  &   1   &       $\;${\bf 843.4}   &   --  &   --  \\
\hline\hline
\end{tabular}}
\caption{\em Diagnostics for the hierarchical clustering algorithm for the
Italian Test - Reading Comprehension section and Grammar section - and the
Mathematics Test: step of the procedure ($h$), number of groups ($s$), increase
of BIC index with respect to the initial model; in boldface are the first
positive values.}\vspace*{0.5cm}
\label{table10}
\end{table}

From Table~\ref{table11}, which shows the support point estimates
for the two sections of the Italian Test and the Mathematics Test,
it can be also shown that, overall, students' belonging to the
higher latent classes is linked with increasing ability levels.

\begin{table}[!ht]\centering
\vspace*{0.5cm}
\begin{tabular}{l|ccccc}
\hline\hline
      & \multicolumn5c{$c$} \\
     & 1 & 2 & 3 & 4 & 5 \\
\hline
\underline{\em Reading Comprehension} &  &  &  &  & \\
Dimension 1 &   -1.193 &  $\;$0.221 & -0.329 & 1.012   & 2.776 \\
Dimension 2 & -1.404 &  -0.859  & -0.049 &  0.646 &  1.378   \\
\hline
\underline{\em Grammar} &  &  &  &  & \\
Dimension 1     &   -0.334  &   $\;$2.244   &   $\;$2.536   &   $\;$2.948   &   4.363   \\
Dimension 2 &   -0.853  &   -0.786  &   $\;$0.812   &   $\;$0.935   &   2.807   \\
Dimension 3     &   -0.827  &   -0.554  &   -2.068  &   $\;$0.598   &   2.384   \\
Dimension 4 &   $\;$0.782   &   $\;$1.224   &   $\;$2.012   &   $\;$2.507   &   3.735 \\
Dimension 5 &   -0.616  &   -1.069  &   -0.623  &   -0.056  &   1.364   \\
\hline
\underline{\em Mathematics} &  &  &  &  & \\
Dimension 1 &   0.995 &  1.509 & 2.060 & --   &  -- \\
\hline\hline
\end{tabular}
\caption{\em Support points estimates for the Italian Test - Reading Comprehension section and Grammar section - and the Mathematics Test}
\label{table11}\vspace*{0.5cm}
\end{table}

Indeed, students belonging to class 5 within the two sections of the Italian Test, and to class 3
within the Mathematics Test, tend to have the highest ability level in relation with the involved
dimensions, whereas students' belonging to the first latent class is generally associated with
lower ability levels. These considerations hold for each dimension but for the first dimension of
the Reading Comprehension Section and the third dimension of the Grammar Section - where higher
than expected ability levels are observed in correspondence with middle latent classes - and for
the fifth dimension of the Grammar Section - where the support point estimate observed by the
first latent class is not the lowest.
\section{Conclusions}\label{sec:conc}
The main objective of this paper is to evaluate the dimensionality
of two national Tests employed to assess middle school Italian
students' performance, testing for the assumption of
unidimensionality which characterizes most Item Response Theory
models used to validate assessment data. We also test if the
assumption of absence of Differential Item Functioning (DIF) is
reasonable for these data. The data were collected in 2009 by the
National Institute for the Evaluation of the Education System
(INVALSI) and refer to two assessment Tests - on Italian language
competencies (Reading comprehension, Grammar) and Mathematical
competencies - administered to middle-school students.

We base our analysis on a class of
multidimensional latent class IRT models which allows us to test unidimensionality
by concurrently taking into account the presence of DIF and that the items may have
non-constant item discrimination power.
This class of models is obtained as an extension for (uniform) DIF of the class of
multidimensional two-parameter logistic (2PL) models developed by \cite{bart:07}.
The inclusion of DIF effects has proven opportune as the hypothesis of
absence of these effects was strongly rejected for both Tests here considered.
Moreover, as known, Tests containing items affected by DIF and, thus, functioning
differently for respondents who belong to different groups, may have a reduced validity.
In the context of this study, the soundness of between-group comparisons is trimmed down
by the dependance of students' scores on attributes other than those the scale is intended
to measure, that is students' gender and geographical area.

Concerning the hypothesis of unidimensionality, the advantage of the applied approach with
respect to other approaches is that it can be employed when the items discriminate differently
among subjects. Within the present analysis, relying on a 2PL parameterisation has been
justified by the lack of any prior information on discriminating power of the test items.

To test the assumption of unidimensionality, we compare a
unidimensional model with a multidimensional counterpart with the
same 2PL parameterisation, the same number $k$ of latent classes,
and the same DIF structure, relying on a Wald test statistic.
Subsequently, we cluster items in different unidimensional groups.
The classification algorithm performed under this set-up showed that
the assumption of unidimensionality is not supported by the data for
the Italian Test, while it can be accepted for the Mathematics Test.
Therefore, while summarizing students' performances on the
Mathematics Test through a single score is appropriate, a single
score cannot be sensibly used to describe students' attainment on
the Italian Test (especially on the Grammar section), as the
difference among students' does not depend univocally on a single
ability level.

\appendix
\section*{Appendix 1: EM algorithm for model estimation}
The {\em complete log-likelihood}, on which the EM algorithm is based,
may be expressed as
\begin{equation}\label{eq:comploglik}
\ell^*(\boeta) = \sum_{c=1}^k\sum_{g=1}^h\sum_{\bl y}
n(c,g,\b y) \log[p_g^*(\b y|c)\pi_c],
\end{equation}
which is directly related to the {\em incomplete log-likelihood} defined
in (\ref{eq:loglik_freq}), and where $n(c,g,\b y)$ denotes the number of
subjects providing response configuration $\b y$ and belonging to
latent class $c$ and to group $g$, whereas $p_g^*(\b y|c)$ corresponds
to the conditional probability defined in (\ref{eq:cond_prob}) for a
subject belonging to the $g$-th group.

Usually, $\ell^*(\boeta)$ is much easier to maximize with respect of
$\ell(\boeta)$. However, since the frequencies $n(c,g,\b y)$ are not known,
the EM algorithm alternates the following two steps until convergence in
$\ell(\boeta)$:
\begin{itemize}
\item \underline{E-step.} It consists of computing the expected value of
the complete
log-likelihood $\ell^*(\boeta)$; this is equivalent to substituting each
frequency $m(c,g,\b y)$ with its expected value
$$
\tilde{m}(c,g,\b y)= n(g,\b y)
\frac{p_g(\b y|c)\pi_c}{\sum_hp_g(\b y|h)\pi_h},
$$
under the current value of the parameters.
\item \underline{M-step.} It consists of updating the model parameters
by maximizing the expected value of $\ell^*(\boeta)$. More
precisely, for the weights $\pi_c$ an explicit solution exists which
is given by
$$
\hat{\pi}_c = \frac{\sum_{g=1}^h\tilde{m}(c,g,\b y)}{n},\quad
c=1,\ldots,k.
$$
About the other parameters, since an explicit solution does not exist, an
iterative optimization algorithm of Newton-Raphon type may be used.
The resulting estimates of $\boeta$ are used to update  $\tilde{m}(c,g,\b y)$
at the next E-step.
\end{itemize}

When the algorithm converges, the last value of $\b\eta$, denoted by
$\hat{\boeta}$, corresponds to the maximum of $\ell(\boeta)$ and
then it is taken as the maximum likelihood estimate of this
parameter vector. It is important to highlight that the running time
and, in particular, the detection of a global rather than a local
maximum point crucially depend on the initialization of the EM
algorithm. Therefore, following \cite{bart:07}, we recommend to try
several initializations of this algorithm that may be formulated in
terms of initial expected frequencies $\tilde{m}(c,g,\b y)$. These
frequencies may be obtained by multiplying each observed frequency
$n(g,\b y)$ by a given constant $\alpha_c(\b y)$ depending on the
total score (i.e., the sum of the elements in $\b y$). These
constants must satisfy the obvious constraints $\al_c(\b y)>0$,
$c=1,\ldots,k$, and $\sum_c\al_c(\b y)=1$ for all $\b y$.

\bibliography{biblio}
\bibliographystyle{apalike}

\end{document}